\documentclass[12pt]{article}

\usepackage{amsmath}
\usepackage{amssymb}
\usepackage{graphicx}
\usepackage{cite}
\usepackage{color}

\usepackage[margin=1in]{geometry}
\usepackage{setspace}
\newcommand{\gn}{G}

\begin{document}

\begin{titlepage}

\vspace*{-15mm}   
\begin{flushright}   
\begin{tabular}{r}    
{\tt NU-QG-16}\\   
%month 20yy
\end{tabular}   
\end{flushright}   

\vspace{10mm}   

\begin{center}
{\Large\bf
 Interior of Schwarzschild in semiclassical gravity
}

\vspace{8mm}

\renewcommand{\thefootnote}{\fnsymbol{footnote}}

Yoshinori~Matsuo\footnote[2]{ymatsuo@nagoya-u.jp} 

\renewcommand{\thefootnote}{\arabic{footnote}}
 
\vspace{5mm}   

{\it  
 Department of Physics, Nagoya University, Nagoya 464-8602, Japan
}
  
\vspace{10mm}   

%\today

\vspace{10mm}   

\end{center}

\begin{abstract}
In Einstein gravity, matter with an arbitrarily small density can be a black hole. 
Pressure in the star diverges if size of the star is smaller than 9/8 of the Schwarzschild radius, 
implying the gravitational collapse into a black hole. 
By taking quantum effects of matter, however, pressure is bounded from above, and 
a core with negative energy appears instead. 
Density of matter increases and eventually reaches the cut-off scale 
as size of the star approaches the Schwarzschild radius. 
This result implies that density must be very large as the Planck scale 
if the star is put inside the Schwarzschild radius. 
%or equivalently, inside a black hole. 
\end{abstract}

\vspace{10mm}   

%\begin{center}
% Essay written for the Gravity Research Foundation 2026 Awards \\ for Essays on Gravitation.
%\end{center}

%\baselineskip 18pt   

\end{titlepage}

\newpage

An important feature of black holes is that 
matter with arbitrarily small density can be put inside the Schwarzschild radius. 
Roughly speaking, the mass $M$ is proportional to the cube of the radius of the star $r_s$ 
while the Schwarzschild radius $r_h$ scales linearly with the mass, $r_h = 2 \gn M$, 
where $\gn$ is the Newton constant. 
Thus, a star with an arbitrary density is hidden behind the Schwarzschild radius 
if the size is sufficiently large. 
Moreover, pressure in the star diverges if the radius $r_s$ 
and the mass $M$ satisfies $r_s \leq \frac{9}{4} \gn M$, 
%where $\gn$ is the Newton constant, 
when the star consists of perfect fluid with a density non-increasing outwards \cite{Buchdahl:1959zz}, 
implying that the star cannot be in a static state 
and causes the gravitational collapse into a black hole. 
%This condition is known as the Buchdahl bound \cite{Buchdahl:1959zz}. 

In semiclassical gravity, since quantum states possibly have negative energy, 
density of matter can increase outwards, and then, 
stars with the radius $r_s \leq \frac{9}{4} \gn M$ can exist. 
Quantum effects of matter are negligible in most cases 
but play an important rule under very strong pressure. 
Here, we study the interior of the Schwarzschild spacetime in semiclassical gravity. 
A core with negative energy appears instead of the divergence of pressure, 
and the density of matter outside the core must be much larger 
to give the same mass to the classical case. 
We focus on the incompressible fluid and show that 
the density of matter $\rho$ is so large as the Planck scale $\rho \sim \gn^{-2}$ 
when the radius of the star is almost the same to the Schwarzschild radius, $r_s \simeq r_h$. 
%The radius of the star cannot be smaller than the Schwarzschild radius $r_h = 2 \gn M$, 
%as long as the density of matter $\rho$ is much smaller than the Planck scale, $\rho \ll \gn^{-1}$. 
%Here, we focus on an incompressible fluid and 
%study the interior of the Schwarzschild spacetime in semiclassical gravity. 
%show how buchdahl bound is modified by quantum effects. 

We consider a static solution of a star which consists of a perfect fluid. 
The metric of static and spherically symmetric spacetime in four dimensions is given by 
\begin{equation}
 ds^2 = - f(r) dt^2 + \frac{dr^2}{h(r)} + r^2 d \Omega^2 \ , 
 \label{metric}
\end{equation}
where $d \Omega^2$ is the metric of the two-dimensional sphere. 
The stress-energy tensor of a perfect fluid in a static configuration is given 
in terms of the energy density $\rho$ and pressure $p$ as 
\begin{align}
 T^t{}_t 
 &= - \rho \ , 
&
 T^i{}_i
 &= 
 p \ , 
\end{align}
where the index $i$ labels spatial components, and the other components are zero. 
By solving the Einstein equation, $h(r)$ is calculated as 
\begin{equation}
 h(r) = 1 - \frac{2\gn m(r)}{r} \ , 
\end{equation}
where $m(r)$ is given in terms of the energy density $\rho$ as 
\begin{equation}
 m(r) = 4\pi \int_0^r {r'}^2 dr'\,\rho(r') \ . 
 \label{m}
\end{equation}
%For simplicity, we assume that the outside of the star $r > r_s$ is vacuum, $T_{\mu\nu} = 0$, 
%and then, 
Since the geometry outside the star $r \geq r_s$ is given by the Schwarzschild metric, 
\begin{equation}
 f(r) = h(r) 
 = 
% 1 - \frac{r_h}{r} \ , 
 1 - \frac{2 \gn M}{r} \ , 
\end{equation}
%where $M$ is the total mass of the star. 
%and $m(r)$ for $r\geq r$ agrees with the total mass $m(r_s) = M$. 
%Although the proper volume element as well as the gravitational potential energy 
%are not taken into account in the integration \eqref{m}, 
%Thus, 
$m(r)$ is related to the total mass of the star $M$ as $M = m(r_s)$. 
%by the junction condition with the Schwarzschild solution. 
For incompressible fluids, $\rho = \text{const.}$, 
the total mass is simply given by $M = \frac{4\pi}{3} r_s^3 \rho$. 
Thus, a star with an arbitrarily small density can be put inside the Schwarzschild radius $r_h = 2 \gn M$, 
if size of the star is sufficiently large, namely $r_s > \sqrt{\frac{3}{8\pi \gn \rho}}$. 

In the case of an incompressible fluid, 
the conservation law of the stress-energy tensor is solved to give an expression of the pressure $p$ as 
\begin{equation}
 p(r) = \left(\sqrt{ \frac{f(r_s)}{f(r)}} - 1\right) \rho \ ,  
\end{equation}
where the integration constant is fixed so that pressure vanishes at the surface of the star, $p(r_s) = 0$. 
By solving the Einstein equation, $f(r)$ and $h(r)$ are calculated as 
\cite{Schwarzschild:1916ae}
\begin{align}
 f(r) 
 &= 
% h(r) 
% \left(
%  \frac{3 \sqrt{1-\frac{2\gn M}{r_s}}}{2 \sqrt{1-\frac{2 \gn M r^2}{r_s^3}}} - \frac{1}{2}
% \right)^2 
 \left(
  \frac{3}{2} \sqrt{1-\frac{2\gn M}{r_s}} - \frac{1}{2} \sqrt{1-\frac{2 \gn M r^2}{r_s^3}} 
 \right)^2 
 \ , 
&
 h(r) 
 &= 
 1 - \frac{2 \gn M r^2}{r_s^3} \ . 
\label{classical}
\end{align}
%where we imposed the conditions that the metric is connected to 
%the exterior solution of the Schwarzschild $f(r) = h(r) = 1 - \frac{2\gn M}{r}$ 
%at the surface of the star $r = r_s$ and that there is no conical singularity at the origin, 
%namely, $h(0)=1$. 
If the radius of the star $r_s$ satisfies the inequality $r_s \leq \frac{9}{4}\gn M$, 
$f(r)$ goes to zero at some radius $r \geq 0$, and hence, the pressure $p$ diverges there. 
When the inequality is saturated $r_s = \frac{9}{4}\gn M$, the divergence appears at $r=0$. 
%If the radius of the star $r_s$ saturates the bound $r_s = \frac{9}{4}\gn M$, 
%$f(r) = 0$ at $r = 0$, and hence, the pressure diverges there. 
%For a smaller star, the divergence of pressure appears at non-zero radius, $0<r<r_s$. 

For solutions of the semiclassical Einstein equation, 
such a divergence of pressure cannot appear due to quantum effects. 
We separate the stress-energy tensor into the fluid part and vacuum part as
\begin{equation}
 \langle T_{\mu\nu} \rangle 
 = 
 T^{\text{(fluid)}}_{\mu\nu} 
 + \langle 0 | T_{\mu\nu} | 0 \rangle \ , 
\end{equation}
where $|0\rangle$ is the vacuum state. 
Each part of the stress-energy tensor satisfies the conservation law separately, 
and the fluid part is expected to approximate the stress-energy tensor of classical fluid 
as it describes the energy and pressure of particle excitations from the vacuum. 
Here, we make a rough estimation of the vacuum stress-energy tensor by using the Weyl anomaly. 
For conformal matter, the trace of the stress-energy tensor satisfies the anomaly condition \cite{Duff:1977ay}, 
\begin{align}
 \langle 0 | T^{\mu}{}_\mu | 0 \rangle 
 &= 
 c\,\mathcal F - a\,\mathcal G \ , 
\\
 \mathcal F 
 &= 
 R^{\mu\nu\rho\sigma} R_{\mu\nu\rho\sigma}
 -2 R^{\mu\nu} R_{\mu\nu}
 + \frac{1}{3} R^2 \ , 
&
 \mathcal G
 &= 
 R^{\mu\nu\rho\sigma} R_{\mu\nu\rho\sigma}
 -4 R^{\mu\nu} R_{\mu\nu}
 + R^2 \ , 
\end{align}
where $R_{\mu\nu\rho\sigma}$, $R_{\mu\nu}$ and $R$ are 
the Riemann tensor, Ricci tensor and Ricci scalar, respectively. 
Both of coefficients $a$ and $c$ depend on matter content but are always positive. 
Around the divergence of pressure at $r = r_c$ on the classical interior solution of the Schwarzschild above, 
$\mathcal F$ and $\mathcal G$ behave as 
\begin{align}
 \mathcal F &\propto \frac{1}{(r-r_c)^2} \ , 
 & 
 \mathcal G &\propto \frac{1}{r-r_c} \ ,  
\end{align}
for $r_c > 0$, and 
\begin{align}
 \mathcal F &\propto \frac{1}{r^4} \ , 
 & 
 \mathcal G &\propto \frac{1}{r^2} \ ,  
\end{align}
for $r_c=0$. Thus, $\mathcal F$ in the Weyl anomaly dominates over $\mathcal G$, 
and hence, the trace part of the stress-energy tensor is of $\mathcal O(R^2)$ and positive. 
Assuming that all curvature squared terms have a similar behavior, 
$R^{\mu\nu\rho\sigma} R_{\mu\nu\rho\sigma} \sim R^{\mu\nu} R_{\mu\nu} \sim R^2$, 
the trace part of the Einstein equation roughly behaves as 
\begin{equation}
 - R \sim 8\pi \gn \left(3p + R^2\right) \ , 
\end{equation}
which implies that the pressure $p$ and curvature $R$ are bounded as 
\begin{equation}
 p < - (24\pi \gn)^{-1} R \lesssim \gn^{-2} \ . 
\label{bound}
\end{equation}
Therefore, the interior solution of the semiclassical Einstein equation 
cannot have the divergence of pressure. 
Although matter does not have the conformal symmetry in general, 
it would be expected that the trace part of the stress-energy tensor 
behaves as $T^\mu{}_\mu \sim R^2$ for sufficiently large curvatures, 
and hence, the condition \eqref{bound} would be valid even for non-conformal matter. 
Quantum effects become non-negligible when the pressure becomes of $\mathcal O(R^2)$, 
or equivalently, around $r = r_c$ and inside it, assuming that the pressure is non-increasing outwards. 
There, the pressure and curvature would behave as $p \sim - \gn^{-1} R \sim \gn^{-2}$. 
Here, we call this region the semiclassical core. 

%The quantum vacuum state on the Schwarzschild spacetime has negative energy. 
%The vacuum state is defined as the state with the minimum eigenvalue of a Hamiltonian, 
%which is associated to a time coordinate. 
%Since the time coordinate of a locally flat frame is different from $t$ in the metric \eqref{metric}, 
%a state with zero energy inside the Schwarzschild spacetime is not 
%the vacuum state associated to the time $t$ but an excited state, 
%implying that the vacuum state has negative energy there. 
%Since the vacuum state associated to the time coordinate $t$ in the metric \eqref{metric} 
Assuming that the stress-energy tensor is zero in the flat spacetime near the spatial infinity, 
%the vacuum state is the Boulware vacuum \cite{Boulware:1974dm}, and then, 
the energy density of the vacuum part $\rho^\text{(vac)} = - \langle 0 | T^t{}_t | 0 \rangle$ 
becomes negative at finite $r$ \cite{Boulware:1974dm}. 
%Although the stress-energy tensor of the Boulware vacuum diverges if $f(r) = 0$, 
%the bound of pressure \eqref{bound} implies $f(r)>0$. 
The stress-energy tensor depends on details of matter, 
and hence, we simply assume that the energy density of the vacuum part 
is negative and of the same order to the trace part, namely, $\rho^\text{(vac)} \sim - \gn^{-2}$. 
Quantum effects are still negligible outside the semiclassical core, 
but $h(r)$ depends on the energy density in the semiclassical core through \eqref{m}. 
Taking the negative energy in the core into account, 
the solution \eqref{classical} is replaced by 
\begin{align}
 f(r) 
 &= 
 h(r) 
 \left(
 1 
  - \frac{3\gn(M+A)}{r_s}\sqrt{1-\frac{2\gn M}{r_s}} \int_r^{r_s} \frac{r' dr'}{r_s^2} h^{-3/2}(r') 
 \right) \ , 
\label{f}
\\
 h(r) 
 &= 
 1 - \frac{2 \gn (M+A) r^2}{r_s^3} + \frac{2 \gn A}{r} \ , 
\end{align}
where $M$ is the total mass of the star, and 
\begin{equation}
 A = - 4\pi \int_0^{r_c} r^2 dr\,\rho^\text{(vac)} 
\end{equation}
is the total amount of negative energy in the semiclassical core, 
which is defined as effects of the negative vacuum energy $\rho^\text{(vac)}$ to the mass \eqref{m}. 
Here, we assumed that the density of the fluid part is a constant, $\rho^\text{(fluid)} = \text{const.}$, 
and then, is related to the total mass of the fluid as 
\begin{equation}
 M^\text{(fluid)} = M + A = \frac{4\pi}{3} r_s^3 \, \rho^\text{(fluid)} \ . 
\end{equation}
If size of the semiclassical core is comparable with the star, $r_c = \mathcal O(r_s \gn^0)$, 
the total negative energy is of the Planck scale $A \sim r_s^3\, \gn^{-2}$, 
and then, the classical mass of the fluid must also be of the Planck scale, 
$M^\text{(fluid)} \sim r_s^3\, \gn^{-2}$. 
In other words, size of the semiclassical core must be so small as $r_c \sim r_s^{1/3} \gn^{1/3}$, 
as long as the total mass of the fluid is the same order to the classical case, 
$M^\text{(fluid)} \sim r_s \gn^{-1}$. 
Thus, the semiclassical core can be treated as a point in the semiclassical approximation 
although it has a finite size. 

\begin{figure}[t]
\begin{center}
\includegraphics{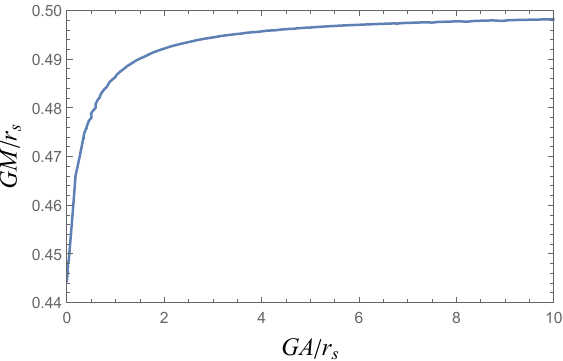}
\caption{
\baselineskip 24pt
As the radius of the star $r_s$ approaches the Schwarzschild radius $r_h = 2 \gn M$, 
the amount of negative energy in the semiclassical core $A$ increases. 
The radius $r_s$ goes to the Schwarzschild radius in $A\to\infty$. 
%implying that the radius of the star cannot be equal to or smaller than the Schwarzschild radius 
%as long as the semiclassical core can be treated as a point. 
}\label{fig}
\end{center}
\end{figure}

The total amount of the negative energy is related to the size of the star $r_s$, 
and the relation is given by the condition that 
the semiclassical core appears around the origin, 
or equivalently, $f(r\sim r_s^{1/3} \gn^{1/3}) \sim \gn^2$. 
This condition can be approximated by $f(0)=0$ 
at the leading order of the small $\gn$ expansion. 
A numerical result of this condition is shown in Fig.~\ref{fig}. 
As the radius of the star $r_s$ approaches the Schwarzschild radius $r_h = 2 \gn M$, 
the amount of negative energy $A$ increases, 
and then, the total mass of the fluid $M^\text{(fluid)} = M + A$ also increases. 
Since the solution \eqref{f} satisfies 
\begin{align}
 f(r) &< h(r) 
 \biggl[
 1 
  - \frac{3\gn(M+A)}{r_s}\sqrt{1-\frac{2\gn M}{r_s}} 
\notag\\
&\qquad\qquad\qquad
 \times \int_r^{r_s} \frac{r' dr'}{r_s^2} 
  \left(
   1 - \frac{6 \gn (M+A)}{r_s} + \frac{2 \gn (2M+3A)}{r} 
  \right)^{-3/2} 
 \biggr]\ , 
\end{align}
the condition $f(0)=0$ leads to 
\begin{equation}
 A > \frac{128 r_s^2}{675\pi^2 \gn (r_s - 2 \gn M)} \ , 
\end{equation}
for $A\gg M$, implying $A\to\infty$ in $r_s \to r_h$. 
The argument above is valid until the negative energy $A$ reaches the Planck scale, 
beyond which the semiclassical core cannot be treated as a point. 
Thus, in contrast to the lower bound of $r_s \geq \frac{9}{8} r_h$ for classical stars, 
size of a star can be approximately equal to but 
slightly larger than the Schwarzschild radius, in semiclassical gravity. 
The negative energy as well as the energy density of matter outside the core 
are so large as the Planck scale 
if the radius of the star is sufficiently close to the Schwarzschild radius. 
%Such a star has a semiclassical core with very large negative energy, 
%and hence, the energy density of matter outside the core is also very large. 
%The radius of the star cannot be smaller than the Schwarzschild radius 
%as long as the semiclassical core can be treated as a point. 
%If the density of the fluid is comparable to the Planck scale, 
%the semiclassical core might have a finite size, 
%and then, the size of the star might be smaller than the Schwarzschild radius. 
%, or equivalently, 
%the density of the fluid is much smaller than the Planck scale $\rho^\text{(fluid)} \ll \gn^{-2}$. 

Our result implies that the structure of spacetime inside the Schwarzschild spacetime 
in semiclassical gravity is quite different from the classical case. 
Matter with a density sufficiently smaller than the Planck scale 
cannot be put inside the Schwarzschild radius 
because the effective volume inside the Schwarzschild radius is too small. 
The proper volume of the star is given by 
\begin{equation}
 V 
 = 
 4\pi \int_0^{r_s} h^{-1/2} r^2 dr 
 \sim 
 r_s^{7/2} \gn^{-1/2} A^{-1/2} 
 \ , 
\end{equation}
if the total amount of negative energy is very large, $A\gg M$. 
It becomes smaller as size of the star becomes smaller. 
A star turns into a black hole when matter is put inside the Schwarzschild radius. 
Then, the energy density of the classical part, or equivalently, particle excitation 
becomes of order of the Planck scale. 
Although our argument does not explain the interior of black holes directly, 
it implies that %the negative energy of the semiclassical core 
%should also be of the same order, $A\sim r_s^3\,\gn^{-2}$, or more, and 
the proper volume inside the black hole would also be so small as $V \sim \gn^{1/2}$ or less, 
in a similar fashion to a semiclassical star slightly larger than the Schwarzschild radius, 
which has negative energy of the Planck scale, $A\sim r_c^3 \gn^{-2}$. 
A black hole would not be a bottomless hole in spacetime 
but its interior would be very small and crammed with collapsed matter. 

Finally, our result here is consistent with 
a numerical study of semiclassical black holes 
by using the s-wave approximation \cite{Ho:2017vgi}, 
in which the semiclassical Einstein equation is solved numerically 
by using the vacuum stress-energy tensor of two-dimensional matter \cite{Davies:1976ei}. 
Although the s-wave approximation is not very good near the center of the star, 
it was found that the energy density of fluid is of order of the Planck scale $\rho \sim \gn^{-2}$ 
if size of the star is approximately equal to the Schwarzschild radius.

\subsection*{Acknowledgments}

This work was supported in part by JSPS KAKENHI Grant No.~JP21H05186.

\end{document}